\documentclass{article}
\usepackage{cite}
\usepackage[dvips]{graphicx}
\newcommand{\be}{\begin{equation}}
\newcommand{\ee}{\end{equation}}
\newcommand{\beqn}{\begin{eqnarray}}
\newcommand{\eeqn}{\end{eqnarray}}
\newcommand{\beqnn}{\begin{eqnarray*}}
\newcommand{\eeqnn}{\end{eqnarray*}}

\def\vep{\varepsilon}

\begin{document}
%\draft

\title{
Photon generation from vacuum in nondegenerate cavities
with regular and random periodic displacements of boundaries
}
\author{ A. V. Dodonov$^{1}$,
E. V. Dodonov$^{2}$ and
V. V. Dodonov$^{3}$
\thanks{corresponding author}
\thanks{ e-mail: vdodonov@fis.unb.br}
\\
%\address{
$^{1}$ Departamento de F\'{\i}sica,
Universidade Federal de S\~ao Carlos, Brasil\\
$^{2}$ Departamento de Computa\c{c}\~ao,
Universidade Federal de S\~ao Carlos, Brasil\\
$^{3}$ Instituto de F\'{\i}sica, Universidade de Bras\'{\i}lia, \\
Caixa Postal 04455, 70910-900 Bras\'{\i}lia, DF, Brasil
}
\date{}
\maketitle

%\small

\begin{abstract}
        We study the influence of fluctuations in
periodic motion of boundaries of an ideal three-dimensional
cavity on the rate of photon generation from vacuum due to
the nonstationary Casimir effect.

\end{abstract}

\vspace{5mm}

{\it PACS}: {42.50.Lc; 46.40.Ff; 72.80.Ng; 03.65.-w}

{\it Key words\/}: 
Dynamical Casimir effect; Periodically moving boundary;
Fluctuations; Transfer matrix; Disordered chain

%\newpage %\twocolumn
%\setlength{\baselineskip}{18pt}
%\renewcommand{\baselinestretch}{0.5}

%\narrowtext
%\twocolumn
%\begin{multicols}{2}

%\normalbaselines

\section{Introduction}

Among many interesting classical and quantum phenomena in cavities
with moving boundaries,
the effect of photon creation from the vacuum
seems to be the most impressive.
The first rough (although not always quite correct) estimations were made
by several authors in Refs.
\cite{Ask62,Riv79,Sarkar,DKM90,Sch,Sas}.
More detailed and consistent studies performed for the past decade
in the frameworks of different approaches
\cite{DK92,DKN93a,BE,Coop,Law,Ditt,D95,Cole,Mep,Lamb,Mund,D98a,D98,Plun,%
Dal99,Llave99,Jing00,AD00,Croc1,Alex,Pluntemp,Plun02,Croc2,Dit02,Petr03}
(for other references see review \cite{review})
predicted, indeed,
an exponential growth of the field energy
in the case of periodic motion and under the
resonance conditions, when the walls vibrate at the frequency
which is a multiple of the unperturbed field eigenfrequency.

Now, an experimental verification of theoretical predictions
becomes a difficult but realizable task.
An idea which seems to be the most realistic at the moment is to
realize an
{\em effective\/} moving mirror by means of periodic creation of a conducting
layer from a semiconductor film posed on the surface of the cavity wall and
irradiated by powerful laser impulses \cite{Carug}.
Different prototypes of this idea were discussed in \cite{Yab,Loz},
but not in a periodic regime
(for other schemes see, e.g., \cite{Kul,Cir99}; however, their realizations
seem to be more difficult).
The effective mirror scheme has several advantages over the primary idea of
an acoustic excitation of harmonical surface vibrations of the wall
\cite{D95}.

The main of them is the removal of severe limitations on the maximal
dimensionless amplitude of the displacement of the surface, which resulted
from tremendous internal stresses inside
the wall whose surface vibrates at the necessary frequency having an order of
several GHz (or higher). In the effective mirror scheme, the maximal
displacement of the boundary depends of the thickness of the semiconductor
film and the laser power, and it can be made several orders of magnitude
larger than the maximal displacements
(not exceeding $10^{-8}\,$cm \cite{D95})
which could be achived in the scheme based on the acoustic
excitation of the surface vibrations.
As a consequence, one can diminish (in the same proportion) the number of
oscillations of the boundary necessary to produce photons in the cavity,
thus relaxing the requirements for the $Q$-factor of the cavity and
the admissible detuning from the exact resonance \cite{D98a,D98}.
Another advantage is a possibility to use
{\em periodic pulses of arbitrary shape\/}, whose period
can be much longer than the period of oscillations of the
chosen mode of electromagnetic field.

However, a practical realization of the scheme depends on its robustness
against inevitable experimental imperfections, such as, e.g., nonexact
periodicity of laser pulses (creating free carriers in the semiconductor
layer) or fluctuations in their power.
Therefore, the aim of our article is to give some estimations of
possible deviations from ideal results caused by such imperfections.

\section{Basic relations}
%Generation of photons %in a single mode
%due to periodic variation of effective cavity length}

The most part of theoretical studies on the nonstationary (dynamical)
Casimir effect was performed within the framework of the model
of a one-dimensional (Fabry--Perot) cavity with {\em equidistant\/}
spectrum of unperturbed eigenfrequencies of the field, when all modes
are coupled in the resonance case. However, in realistic
three-dimensional cavities the spectrum is, as a rule, nonequidistant.
This results in great simplifications in the resonant case, when
periodic motion of the boundary excites only one resonant mode,
whereas the response of all other modes can be neglected in the long-time
limit \cite{D95}. Therefore, we confine ourselves here to this
simplest case (interesting phenomena which could occur in cavities with
few accidentally resonant modes, when the ratio of their unperturbed
eigenfrequencies is integral number due to some additional symmetry,
were considered recently in \cite{Croc1,Alex,Croc2}).

In this case, as was shown in \cite{D95}, the problem of photon
generation in the selected mode is reduced to the problem of excitation
of a quantum oscillator with a time dependent frequency $\omega(t)$,
which is determined by the instantaneous geometry of the
cavity. For example, in a rectangular cavity with fixed dimensions
$L_x,L_y$ and a variable dimension $L_z(t)$ we have
\be
\omega(t)=\pi c\left[\frac{n_x^2}{L_x^2} +
\frac{n_y^2}{L_y^2} +\frac{n_z^2}{L_z^2(t)}\right]^{1/2}.
\label{freq}
\ee
For fifty years passed after the seminal paper by Husimi \cite{Hus},
this last problem was studied in detail in numerous publications
(for reviews see, e.g., \cite{183-2,book}).
It appears that all
properties of the {\em quantum\/} oscillator are determined,
as a matter of fact,
by the fundamental set of solutions of the {\em classical\/} equation
of motion
\be
\ddot{x} +\omega^2(t) x=0.
\label{eq}
\ee
Here we actually need only one consequence of the general theory,
namely, that the mean number of quanta generated from the initial
oscillator ground state due to the time dependence of frequency
is given by the {\em energy reflection coefficient\/} from an effective
``potential barrier'' given by the function $\omega^2(t)$.
More precisely, if $\omega(t) = \omega_i $ for $t\to - \infty$
and $\omega(t) = \omega_f $ for $t\to  \infty$
(this means that the cavity wall is supposed to assume some fixed
positions before and after the experiment), then one has to calculate
the coefficients of the
asymptotical form of the solutions to Eq. (\ref{eq}),
\be
x(t\to \infty) =\omega_f^{-1/2}\left[\xi e^{i\omega_f t} +
\eta e^{-i\omega_f t}\right],
\label{asgen}
\ee
satisfying the initial condition
$x(t) = \omega_i^{-1/2} e^{i\omega_i t}$ for $t\to -\infty$.
Due to the unitarity of evolution, these coefficients obey the identity
\be
|\xi|^2 -|\eta|^2=1.
\label{iden}
\ee
The mean number of quanta for $t\to\infty$ can be expressed in terms of
$\xi$ and $\eta$ as follows \cite{Sas,DKN93},
\be
{\cal N}= |\eta|^2={R}/{T},
\label{Ngen}
\ee
where
$R\equiv |{\eta}/{\xi}|^2$ and $T\equiv 1-R\equiv |\xi|^{-2}$
can be interpreted as energy reflection and transmission
coefficients from the effective potential barrier.

Eq. (\ref{Ngen}) shows that no photons can be created during a single
pass of the wall from one position to another,
because for {\em monotonous\/} functions $\omega(t)$
the reflection coefficient is limited by the Fresnel formula
\be
R_1=\left(\frac{\omega_i -\omega_f}{\omega_i +\omega_f}\right)^2
\label{RF}
\ee
corresponding to very rapid change of position (for the time much less
than the period of field oscillations) \cite{Viss}.
Since real boundaries in laboratory
experiments can only move with velocities much less than the speed of light,
the process is almost adiabatic, and $R_1\ll 1$.
However, it is well known that the reflection coefficient can be made
very close to unity in the case of {\em periodic\/} variations of
parameters due to interference effects.

\begin{figure}[htb]
  \begin{center}
\includegraphics[scale=0.6]{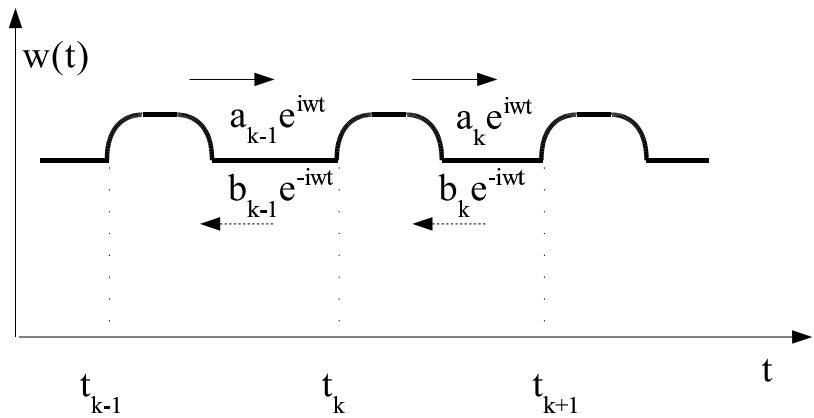}
\vspace{-3in}
\caption{
A typical time dependence of the effective frequency.
}
\label{fig-scheme}
  \end{center}
\end{figure}

For a periodic motion of the wall, the function $\omega^2(t)$ is also
periodic. We assume that it has a form like that shown in
Fig.~\ref{fig-scheme}. The most
important assumption is that each ``effective potential barrier'' is
well separated from the next one by some interval of time where
$\omega=const$ (i.e., that the wall moves for some period of time,
returns to its initial position, stays at this position for some time,
and then repeats the cycle). Thus we exclude monochromatic oscillations
of the boundary, which have been already studied in detail in
\cite{D95}.
%\end{multicols}
%\begin{multicols}{2}
In such a case, each ``barrier'' can be completely characterized by
two complex amplitude reflection coefficients and two complex
amplitude transmission coefficients, which connect the ``plane waves''
coming from the ``left'' and from the ``right''. Namely, if some
``barrier'' begins at $t=0$ at terminates at $t=t_*$, then one can write
two independent solutions of Eq. (\ref{eq}) outside the ``barrier'' as
(hereafter we consider for simplicity the case of equal initial and
final constant values of the frequency $\omega$)
\beqn
&& x^{(-)}(t)= \left\{
\begin{array}{ll}
e^{i\omega t} +r^{(-)}e^{-i\omega t}, & t<0 \\
t^{(-)}e^{i\omega t}, & t>t_*
\end{array} \right.
\label{x-}
\\
&& x^{(+)}(t)= \left\{
\begin{array}{ll}
t^{(+)}e^{-i\omega t}, & t<0 \\
e^{-i\omega t} +r^{(+)}e^{i\omega t}, & t>t_*
\end{array} \right.
\label{x+}
\eeqn
(We use the letter $t$ without supescripts to denote the time variable,
whereas the same letter supplied with supescripts means the amplitude
transmission coefficient; we hope this will not lead to a confusion.)
The coefficients $r^{(\pm)}$ and $t^{(\pm)}$ are not independent,
because equation (\ref{eq}) is invariant with respect to complex
conjugation (the frequency $\omega(t)$ is real). The following
relations hold (see, e.g., \cite{Fadd,Lamboo};
the simplest way to obtain these
relations is to calculate Wronskians for suitable pairs of independent
solutions):
\beqn
&&t^{(-)} = t^{(+)} \equiv t^{(\pm)},
\label{f+-} \\
&& r^{(-)} t^{(\pm)*} + r^{(+)*}t^{(\pm)} =0,
\label{rf+-} \\
&& |r^{(-)}|^2 + |t^{(-)}|^2 = |r^{(+)}|^2 + |t^{(+)}|^2 =1.
\label{r2f2}
\eeqn

Suppose now that we have several well separated ``barriers'',
the $n$th ``barrier''
(possessing coefficients $r_{n}^{(\pm)}$ and $t_{n}^{(\pm)}$ if it begins
at $t=0$) being shifted in time by the value $ t_{n-1}$ (with $t_0=0$)
with respect to the beginning of the first one.
Then one can combine
the first $n$ barriers in a single effective barrier, characterized by
the amplitude reflection and transmission coefficients $\rho_n^{(\pm)}$
and $\tau_n^{(\pm)}$.
There are many different methods enabling to express these
coefficients in terms of $r_{k}^{(\pm)}$ and $t_{k}^{(\pm)}$,
$k=1,\ldots,n$: see, e.g., \cite{Brekh,Stol} and references therein.
We use the method of {\em transfer matrix\/}
\cite{Sas,Wein47,BW,Yeh,Erd82,DMM93,Band93,Fer}
(see also \cite{Grif01} for historical review and
\cite{Long00,Geor01,Monz02} for the most recent applications).

Let us write the solutions to Eq. (\ref{eq}) after the
$k$th barrier (in the region where $\omega(t)=\omega=const$) as
\be
x^{(k)}(t)= a_k e^{i\omega t} + b_k e^{-i\omega t},
\label{solAB}
\ee
with suffix $0$ related to the solution before the first barrier.
Evidently, every two sets of the nearest constant coefficients,
$(a_{k-1},b_{k-1})$ and $(a_{k},b_{k})$, are related by means of a linear
transformation
\be
\left(
\begin{array}{c}
a_{k} \\
b_{k}
\end{array}
\right)
= M_k
\left(
\begin{array}{c}
a_{k-1} \\
b_{k-1}
\end{array}
\right).
\label{abM}
\ee
Comparing Eqs. (\ref{solAB}) and (\ref{abM}) with (\ref{x-}) and (\ref{x+}),
and taking into account the identities
(\ref{f+-})-(\ref{r2f2}), one can express the elements of matrix $M_k$
as follows:
\be
M_k= \left\Vert
\begin{array}{cc}
f_{k}^* & -g_k^* \\
-g_k & f_k
\end{array}
\right\Vert,
\label{Mk}
\ee
where
\be
f_k\equiv [t_k^{(\pm)}]^{-1}, \quad
g_k\equiv r_k^{(-)}/t_k^{(\pm)}.
\label{fg-rk}
\ee
Consequently, each matrix $M_k$ is {\em unimodular\/}:
\be
\det M_k = |f_k|^2 - |g_k|^2 \equiv 1.
\label{unimod}
\ee

We shall use the notation
\be
{\cal M}_n= \left\Vert
\begin{array}{cc}
F_{n}^* & -G_n^* \\
-G_n & F_n
\end{array}
\right\Vert
\label{Mcaln}
\ee
for the total transfer matrix of $n$ barriers, shifted in time by
$t_k$, $k=1,\ldots,n-1$, with respect to the initial instant $t=0$.
Applying consecutively the transformations (\ref{abM}) with account of the
phase shifts $\theta_k \equiv \omega t_k$ (with $\theta_0=0$),
one can obtain the following matrix formula:
\be
{\cal M}_n = \Phi_{n\!-\!1} M_n \Phi_{n\!-\!1}^{\dagger}
\Phi_{n\!-\!2} M_{n\!-\!1}
\cdots \Phi_{1} M_2 \Phi_{1}^{\dagger} M_{1},
\label{Mntot}
\ee
where
\be
\Phi_k \equiv \left\Vert
\begin{array}{cc}
\exp(-i\theta_k) & 0 \\
0 & \exp(i\theta_k)
\end{array}
\right\Vert.
\label{Phik}
\ee
Eq. (\ref{Mntot}) is equivalent to the recursive relations
\be
{\cal M}_k = \Phi_{k-1} M_k \Phi_{k-1}^{\dagger} {\cal M}_{k-1},
\label{recMcal}
\ee
\be
F_k = g_k G_{k-1}^* e^{2i\theta_{k-1}} + f_k F_{k-1},
\label{recF}
\ee
\be
G_k = g_k F_{k-1}^* e^{2i\theta_{k-1}} + f_k G_{k-1},
\label{recG}
\ee
where $F_0=1$ and $G_0=0$. One can verify that Eqs.
(\ref{recF}) and (\ref{recG}) preserve the identities
\be
|F_k|^2 -|G_k|^2 \equiv 1.
\label{idenFG}
\ee
The total mean number of quanta (\ref{Ngen}) created after $n$ impulses
is nothing but
\be
{\cal N}_n \equiv |G_n|^2.
\label{N-G}
\ee
The coefficients $\rho_n$ and $\tau_n$ themselves obey the
{\em nonlinear\/} recurrence relations
\be
\rho_n = \frac{\rho_{n-1} +r_n s_{n-1} \exp(2i\theta_{n-1})}
{1 + \rho_{n-1}^* r_n s_{n-1} \exp(2i\theta_{n-1})},
\label{recrho}
\ee
\be
\tau_n = \frac{\tau_{n-1} t_n^{(\pm)} }
{1 + \rho_{n-1}^* r_n s_{n-1} \exp(2i\theta_{n-1})},
\label{rectau}
\ee
where
$s_n \equiv {\tau_n}/{\tau_n^*}$, $ \rho_1=r_1^{(-)}$,
$ \tau_1=t_1^{(\pm)}$.

\section{Strictly periodic motion of boundaries}\label{sec-strict}

Formula (\ref{Mntot}) becomes especially useful in the case of
{\em strictly periodic motion of boundaries\/}, when all one-barrier
matrices $M_k$ coincide with $M_1$, and $\Phi_k \equiv \Phi_1^k$:
\be
{\cal M}_n = \Phi_{1}^n (\Phi_{1}^{\dagger} M_{1})^n.
\label{Mnper}
\ee
Obviously, the matrix $\Phi_{1}^{\dagger} M_{1}$ is unimodular, as well as
matrix $M_{1}$. Therefore, we can use the well-known formula for the
powers of any two-dimensional unimodular matrix $S$
(see, e.g., \cite{BW,Grif01}):
\be
S^n= U_{n-1}(z)S -U_{n-2}(z)E, \quad z\equiv \frac12\mbox{Tr}S,
\label{Sn}
\ee
where $E$ means the unit matrix and $U_n(z)$ is the Tchebyshev polynomial
of the second kind.

In the case discussed, introducing the new parameter $\nu$
according to the relations
(hereafter $\theta \equiv \theta_1$, $g\equiv g_1$, $f\equiv f_1$)
\be
z= \mbox{Re}\left(f e^{-i\theta}\right) = \pm \cosh\nu,
\label{defnuz}
\ee
and using the hyperbolic representation of the Tchebyshev polynomial
\be
U_n(\cosh\nu)\equiv \frac{\sinh[(n+1)\nu]}{\sinh(\nu)},
\label{Uncosh}
\ee
we arrive at the following explicit expressions for the elements of
matrix ${\cal M}_n$ (for simplicity, we suppose that the sign of variable
$z$ in Eq. (\ref{defnuz}) is positive; this sign does not affect the
final result (\ref{numper}) for the number of created photons):
\be
G_n =g\,\frac{\sinh(n\nu)}{\sinh(\nu)} e^{i\theta(n-1)},
\label{Gnper}
\ee
\be
F_n =f\,\frac{\sinh(n\nu)}{\sinh(\nu)} e^{i\theta(n-1)}
-\frac{\sinh[(n-1)\nu]}{\sinh(\nu)} e^{i\theta n}.
\label{Fnper}
\ee

Consequently, the number of created photons after $n$ impulses equals
\be
{\cal N}_n =|g|^2\frac{\sinh^2(n\nu)}{\sinh^2(\nu)}.
\label{numper}
\ee
The generation of photons is possible provided parameter $\nu$ is real,
i.e.,
\be
|f|\,|\cos(\varphi -\theta)|>1, \quad f\equiv |f| e^{\varphi}.
\label{condf}
\ee
Since $|f|>1$, one has to ajust the phase shift $\theta$ to the phase
of the inverse transmission coefficient $\varphi$. The maximal effect
is achieved for
\be
\theta= \theta_{res} \equiv  \varphi + \pi m
\label{restheta}
\ee
 (with $m$ an integer). This is equivalent to
the following relation between the periodicity of impulses $T=\theta/\omega$,
the period of the electromagnetic field oscillations in the mode concerned
$T_0 =2\pi/\omega$ and the phase $\varphi$:
\be
T=\frac{T_0}{2}\left(m + {\varphi}/{\pi}\right).
\label{cond-T}
\ee
In particular, for nonmonochromatic oscillations of the boundary,
the field mode can be excited not only under the condition of the
parametric resonance $T=T_0/2$, but the period of motion of the
boundary can be greater than the period of field oscillations.
This fact may be important for possible experimental arrangements.

Under the condition (\ref{cond-T}), $\cosh(\nu)=|f|$  and $\sinh(\nu)=|g|$,
so that
\beqn
{\cal N}_n^{(res)} \!&=&\!\! \sinh^2\left(n \nu\right)
\label{N-reg} \\
\!&\equiv&\!\! \frac14 \left[\left(\frac{1\!+\!|r|}{1\!-\!|r|}\right)^{n/2}
-\left(\frac{1\!-\!|r|}{1\!+\!|r|}\right)^{n/2}\right]^2 ,
\nonumber
\eeqn
where $r\equiv r_1^{(-)}$ is the amplitude reflection coefficient
from each barrier
(remember the relations (\ref{fg-rk}) between different coefficients).
If $|r|\ll 1$, then one can replace $\nu$ by $|r|$. Then for $n|r|>1$
we have
\be
{\cal N}_n^{(res)} \approx \frac14 \exp(2n|r|).
\label{N-reg1}
\ee

In the case
of a {\em harmonically oscillating boundary\/} near parametric resonance,
\begin{equation}
\omega(t)=\omega_0\left[1+2\varepsilon\sin(\Omega t)\right],
\quad
\Omega=2(\omega_0 +\delta)
\label{omeg-t}
\end{equation}
($\omega_0$ is the unperturbed field eigenfrequency, and $\Omega$ is
the frequency of the wall vibrations),
the number of quanta created from vacuum is given
by the expression \cite{D98a}
(under the conditions $|\delta|\ll \omega_0$ and $|\varepsilon|\ll 1$)
\be
{\cal N}= \frac{\sinh^2\left(\omega_0\vep \gamma t\right)}{\gamma^2},
\quad \gamma^2=1-\frac{\delta^2}{(\omega_0\vep)^2}\,.
\label{old}
\ee
For $\delta=0$ (exact resonance \cite{D95}), (\ref{old}) coincides
with (\ref{N-reg}), if one identifies $\nu_{ef}=\pi\vep$ (the number of full
cycles equals $n=\Omega t/(2\pi)= \omega_0 t/\pi$).

We see that the concrete form of the law of motion of the boundary $L(t)$
turns out to be not very important
for the exponential dependence of the number
of created quanta on the number of wall's oscillations $n$
for $n \nu \gg 1$
(although it influences the concrete value of the
coefficient $\nu$).
What is important, it is the
fulfillment of the condition (\ref{cond-T}).

The generation of quanta in the harmonic case becomes impossible
if $\delta > \omega_0\vep$. This means that the phase deviation
from the resonance value accumulated for one cycle,
$\Delta\theta =\delta\cdot T = \delta\cdot\pi/\omega_0$,
should not exceed the critical value
$\Delta\theta_c=\pi\vep =\nu_{ef}$.
The same result holds, in fact, in the general nonmonochromatic case.
Indeed, for small values of $|g|\ll 1$ we have
$|f|=\sqrt{1+|g|^2}\approx 1+ |g|^2/2$.
Then Eq. (\ref{defnuz}) results in
$\nu^2 \approx |g|^2 - |\Delta\theta|^2$,
where $\Delta\theta \equiv \theta -\theta_{res}$
(see Eq. (\ref{restheta}) for the definition of the resonance phase
$\theta_{res}$),
so that
\be
{\cal N}_n \approx \frac{|g|^2}{|g|^2 - |\Delta\theta|^2}
\sinh^2\left(n\sqrt{|g|^2 - |\Delta\theta|^2}\right).
\label{numperdel}
\ee
The critical regular phase shift per cycle equals
$|\Delta\theta_c|=|g|=\sinh(\nu_{res})\approx \nu_{res}$.
By the order of magnitude, taking into account the Fresnel limit
(\ref{RF}), we can estimate the maximal phase deviation per cycle
(from the resonance value)
as $|\Delta\theta_c|<\Delta L/(2L)$, where $\Delta L$ is the change
of the cavity length.

\section{Influence of irregularities of periodic motion}

In real experiments, it can be difficult to maintain the resonance
conditions $g_k \equiv g_1$, $f_k \equiv f_1$, $\theta_k =k\theta_{res}$
exactly. Therefore, it is important to evaluate the influence
of systematic and random deviations (fluctuations) from these conditions.

\subsection{Amplitude variations}
It seems that
fluctuations of
absolute values of the effective reflection and transmission coefficients,
or coefficients $|g_k|$ and $|f_k|$,
are less important than
fluctuations of phases $\theta_k$
(i.e., unequal intervals between pulses).
To show this, let us notice that
for identical barriers,
the resonance condition (\ref{restheta}) is equivalent to the requirement
that the second term in the
denominators of fractions in Eqs. (\ref{recrho}) and (\ref{rectau})
is {\em real\/}. In such a case, all coefficients $\rho_n$ have
{\em identical\/} phases. Moreover, one can adjust the moments of time
$t_k$ in such a way that this property holds even for barriers of
different heights. Under such ``generalized resonance conditions'',
\be
\zeta_{n+1} - \zeta_{n} -  2\varphi_{n} +2(\theta_{n+1} -\theta_{n})
= 0 \;\;\; \mbox{mod}(2\pi),
\label{genrescond}
\ee
where $r_n=|r_n|\exp(i\zeta_n)$ and $f_n=|f_n|\exp(i\varphi_n)$,
Eq. (\ref{recrho}) assumes the form
\be
|\rho_n| = \frac{|\rho_{n-1}| +|r_n|}
{1 + |\rho_{n-1}| |r_n|}.
\label{recrhoabs}
\ee
Writing $|r_n|=\tanh(\nu_n)$, one can verify that
\be
|\rho_n|=\tanh\left(\sum_{k=1}^n \nu_k\right),
\label{solabs}
\ee
and the number of quanta in the selected mode monotonously increases with
time as
\be
{\cal N}= \sinh^2\left(\sum_{k=1}^n \nu_k\right).
\label{n-id}
\ee
For strictly periodic processes ($\nu_n=\nu=const$)
formula (\ref{n-id}) coincides with (\ref{N-reg}).

\subsection{Phase fluctuations}

Now we suppose that reflection and transmission coefficients are
exactly the same for all barriers,
but the strict periodicity is broken:
\be
\theta_k =k\theta +\chi_k, \qquad
f_k \equiv f, \quad g_k \equiv g,
\label{stoch}
\ee
where $\chi_k$ is a stochastic variable.
To have a qualitative understanding, what can happen in this situation,
let us consider first the case, where $\chi_k$ is in fact regular function,
namely,
\[
\chi_{2k+1}=+\chi, \quad \chi_{2k}=-\chi,
\]
which nonetheless destroys a strict periodicity of pulses.
This problem still can be solved analytically,
because each two consecutive barriers can be combined in new effective
barriers, forming strictly periodic sequence.
The total transfer matrix after $2n$ pulses can be represented
in the following form:
\[
{\cal M}_{2n} = \Phi_{1}^{2n-1}X\left(M' X^{-2}M' X^2\right)^{n-1}
M' X^{-1}M' \Phi_{1}
%\label{M2n}
\]
where
\[
M' = M_1 \Phi_{1}^{-1}= \left\Vert
\begin{array}{cc}
\tilde{f}^* \! &\! -\tilde{g}^* \\
-\tilde{g} \!&\! \tilde{f}
\end{array}
\right\Vert,
\;
X= \left\Vert
\begin{array}{cc}
e^{-i\chi} \!&\! 0 \\
0 \!&\! e^{i\chi}
\end{array}
\right\Vert
\]
and $\tilde{f}=f e^{-i\chi}$, $\tilde{g}=g e^{i\chi}$.
Using again formula (\ref{Sn}), we can express elements of the total
transfer matrix in terms of the Tchebyshev polynomials of the argument
$z=\mbox{Re}(\tilde{f}^2) +|g|^2\cos(4\chi)$. We omit intermediate
calculations and bring here the final simplified result,
which is valid for $|g|\ll 1$, $n \gg 1$ and $|\Delta\theta| \ll 1$,
Under these restrictions (which correspond to expected experimental
conditions), we can neglect the difference between $U_n(z)$ and
$U_{n-1}(z)$ for $z$ close to $1$ and replace $\cos(2\Delta\theta)$
by $1-2(\Delta\theta)^2$. Moreover, we can neglect the diference between
the values ${\cal N}_{2n}$ and ${\cal N}_{2n-1}$.
Finally, we arrive at the following simple formula:
\be
{\cal N}_{n} \approx \frac{\cos^2(2\chi)
\sinh^2\left(|g|n\sqrt{
\cos^2(2\chi) - (\Delta\theta/|g|)^2}\right)}
{\cos^2(2\chi) - (\Delta\theta/|g|)^2 }\,.
\label{N-chi}
\ee
We see that the presence of phase ``jumps'' diminishes the rate
of increase of the number of photons, although the growth remains exponential
for sufficiently large number of pulses.
The admissible level of jumps is determined by the inequality
$|\cos(2\chi)| > |\Delta\theta|/|g|$, which shows that ``random''
deviations from the resonance are much less dangerous than systematic ones.
Indeed, the maximal admissible value of the systematic phase shift
per period equals
$\Delta\theta_{max}=|g|$, in complete acordance with evaluations made at the
end of  section \ref{sec-strict}.
On the other hand, for zero
systematic shift from the exact resonance ($\Delta\theta=0$),
exponential growth of the number of photons can be observed
for any amplitude of ``phase jumps'' $\chi$, although the
growth becomes slower with increase of $\chi$ in the interval $(0,\pi/4)$:
\be
{\cal N}_{n} \approx
\sinh^2\left(|g|n |\cos(2\chi)| \right).
\label{N-chi0}
\ee

\begin{figure}[htb]
  \begin{center}
\includegraphics[scale=0.3,angle=-90]{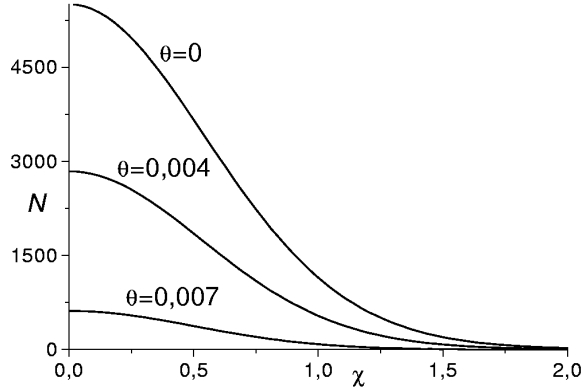}
\caption{
Mean number of photons after $n=500$ pulses
(with $|g|=0.01$)
versus the maximal amplitude of random
phase fluctuations $\chi$ for fixed values of systematic phase deviations
$\theta$.
}
\label{fig-chi}
  \end{center}
\end{figure}

\begin{figure}[hbt]
  \begin{center}
\includegraphics[scale=0.3,angle=-90]{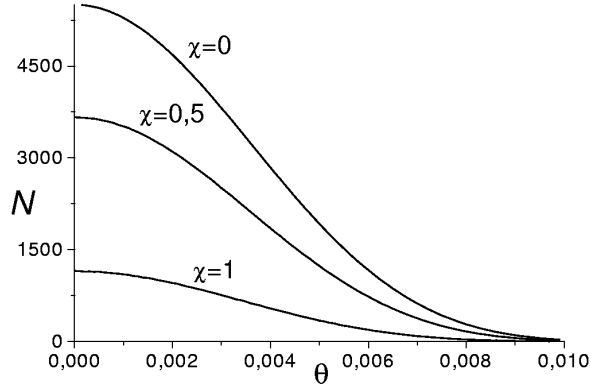}
%\vspace{2.5in}
\caption{
Mean number of photons after $n=500$ pulses
(with $|g|=0.01$)
versus the systematic phase deviations $\theta$,
for fixed values of the amplitude of phase fluctuations $\chi$.
}
\label{fig-theta}
  \end{center}
\end{figure}

\begin{figure}[htb]
  \begin{center}
\includegraphics[scale=0.33,angle=-90]{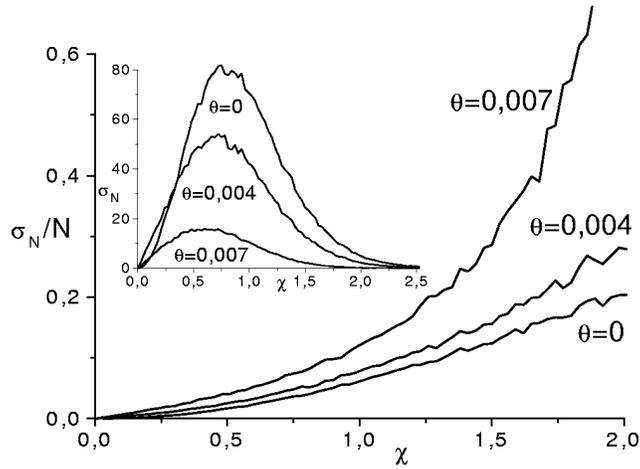}
%\vspace{2.5in}
\caption{
The relative mean-squared deviation of the number of created photons
versus the maximal amplitude of random
phase fluctuations $\chi$ for fixed values of systematic phase deviations
$\theta$ (for $n=500$ pulses).
The absolute mean-squared deviations are shown in the insertion.
}
\label{fig-var}
  \end{center}
\end{figure}

\begin{figure}[hbt]
  \begin{center}
\includegraphics[scale=0.3,angle=-90]{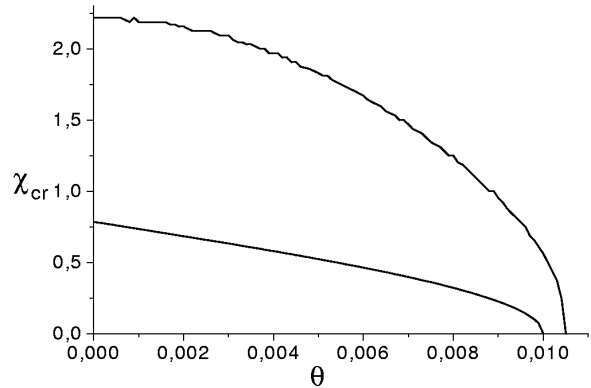}
%\vspace{2.5in}
\caption{
The dependence of the critical amplitude of random
phase fluctuations
on the systematic phase shift $\theta$ (the upper curve),
compared with the maximal amplitude of ``phase jumps'',
when the photon generation stops
according to Eq. (\ref{N-chi})
(the lower curve).
}
\label{fig-crit}
  \end{center}
\end{figure}

We have performed numerical experiments on the basis of recursive relations
(\ref{recF}) and (\ref{recG}), choosing parameters $g$ and
$f=\sqrt{1+g^2}$ in Eq. (\ref{stoch}) real
(so that the resonance condition is
$\theta_{res}=\pi m$; obviously, this choice does not affect the
qualitative picture), namely, we chose $g=0.01$ (in order to reduce the time
of calculations and make the results more vizualizable).
The coefficients $\chi_k$ were chosen with the aid of a program
generating random numbers in the interval $(-\chi,\chi)$, for different
maximal values $\chi$ and systematic displacements $\theta=\Delta\theta$ from
$\theta_{res}=0$ (which is equivalent to $2\pi m$, of course).
The results are given in the figures \ref{fig-chi}-\ref{fig-crit}.

Fig.~\ref{fig-chi} shows the mean number of photons
in the cavity after a fixed number of $n=500$ pulses
(which generate
approximately $\sinh^2(5)\approx 5500 $ photons for $g=0.01$
in the case of strict resonance)
versus the maximal amplitude of random
phase fluctuations $\chi$ for fixed values of systematic phase deviations
$\theta$. The averaging was performed in all cases over $700$ realizations
consisting of $500$ random numbers $\chi_k$.
The average values
$\langle \chi_k \rangle =n^{-1} \sum_{k=1}^{n} \chi_k$
varied from
$10^{-7}$ to $10^{-5}$ for each realization (so that no systematic
shift was added due to random choice of $\chi_k$).
Fig.~\ref{fig-theta} gives the mean number of photons
versus $\theta$ for different fixed values of $\chi$.
Fig.~\ref{fig-var} shows the relative (and in the insertion -- absolute)
mean-squared deviations from the average numbers, as functions of $\chi$
for $\theta=0$.
In Fig.~\ref{fig-crit}, the upper curve shows the dependence of the
critical amplitude of random
phase fluctuations (defined by the condition that the average
number of created photons after $n=500$ pulses is between $9$ and $10$)
on the regular phase shift $\theta$; the lower curve gives the
maximal amplitude of phase ``jumps'', when the photon generation stops
according to Eq. (\ref{N-chi}).
It is clearly seen, that even very high level of {\em random\/}
phase fluctuations does not stop the generation of photons, contrary
to {\em systematic\/} phase shifts.
Moreover, the critical amplitude of truely random fluctuations
remains nonzero even for systematic phase shifts exceeding the
value $|g|$.

\subsection{Analogy with resistance of disordered chains}

It is worth noting that the problem of transmission through a sequence of
one-dimensional barriers
with randomly varying parameters (in particular, interbarrier distances)
has been extensively studied in the context of solid state physics
\cite{Erd82}. There, one of objectives is
the resistance ${\cal R}$ of some chain of scatteres,
expressed (in dimensionless units)
by {\em exactly the same\/} formula (\ref{Ngen}) which gives the
mean photon number in our case.
In particular, Landauer \cite{Land70} used
a consequence of Eq. (\ref{recrho}),
\be
\frac{|\rho_n|^2}{1\!-\!|\rho_n|^2} = \frac{|\rho_{n\!-\!1}|^2 +|r_n|^2
+2|r_n \rho_{n\!-\!1}| \cos(\Psi_n)}
{\left(1\!-\!|\rho_{n\!-\!1}|^2\right)\left(1\!-\!|r_n|^2\right)},
\label{recrho2}
\ee
\[
\Psi_n = \zeta_n + 2 \theta_{n-1} + 2 \arg(s_{n-1}) -\arg(\rho_{n-1}),
\]
and, supposing that the phase shift $\Psi_n$ is a random variable,
uniformly distributed over a large interval,
performed averaging over $\Psi_n$  directly
in Eq. (\ref{recrho2}), having replaced
the term with $\cos(\Psi_n)$ by zero.
One can easily verify that the change of variables,
$|\rho_n|^2=\tanh(y_n)$, $|r_n|^2=\tanh(z_n)$,
transformes the
set of recurrence relations (\ref{recrho2}) with $\cos(\Psi_n)$ replaced
by zero, to the equations
$y_n=y_{n-1} +z_n$, whose obvious solution is $y_n=\sum_{k=1}^n z_k$.
Thus we arrive at the following formula for the resistance of
a chain with completely random distances between scatterers, uniformly
distributed over sufficiently large intervals:
\be
{\cal R}_n \equiv {\cal N}_n=
\frac12 \prod_{k=1}^n \frac{1+|r_k|^2}{1-|r_k|^2} -\frac12.
\label{Resnew}
\ee
Here $|r_k|^2$ is the energy reflection coefficient from the $k$th obstacle
(barrier) and $n$ is the number of obstacles.
If $|r_n|^2=|r|^2=const$, then (\ref{Resnew})
goes to the Landauer formula \cite{Land70}
\be
{\cal R}_n \equiv {\cal N}_n=
\frac12\left(\frac{1+|r|^2}{1-|r|^2}\right)^n -\frac12.
\label{Res}
\ee
It is worth comparing Eq. (\ref{Res}) with Eq. (\ref{N-reg}),
paying attention to different powers of the reflection coefficient $|r|$.
For $n|r|^2>1$ formula (\ref{Res}) yields
\be
{\cal N}_n^{(rand)} \approx \frac12 \exp(2n|r|^2).
\label{N-res1}
\ee
Consequently, totally random phase fluctuations {\em without systematic
phase shift\/} must also lead to an exponential growth of the photon number
(chain resistance), although with much more slow rate than in the
resonance case, if $|r|\ll 1$. For example, for $|r|=0.01$
one needs about 500 pulses
to create 5000 photons under the resonance conditions,
according to Eq. (\ref{N-reg1}),
whereas Eq. (\ref{N-res1}) suggests that the same number of photons
can appear after about
$5\cdot10^{4}$ pulses in the case of completely random fluctuations
(with zero systematic phase shift).

\begin{figure}[htb]
  \begin{center}
\includegraphics[scale=0.5,angle=-90]{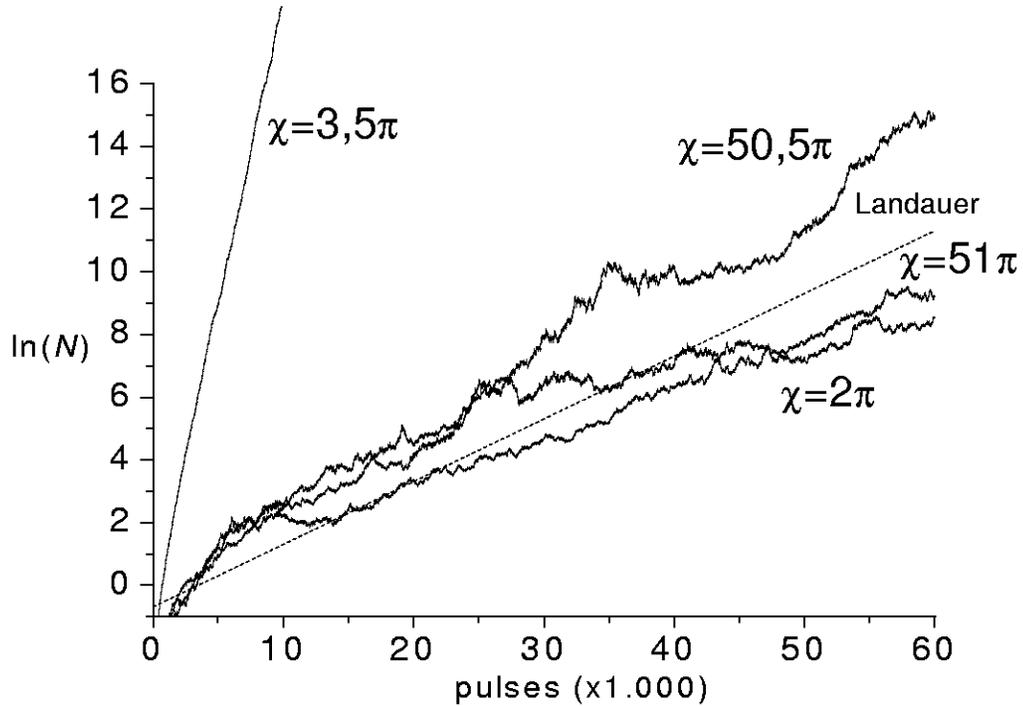}
%\vspace{2.5in}
\caption{
The natural logarithm $\mbox{ln}\langle{\cal N}_n\rangle$
of the average number of photons created after $n$ pulses with
completely random phase $\chi_k$,
uniformly distributed in the interval $(-\chi,\chi)$,
for $\theta=0$, $g=0.01$ and different values of $\chi$.
The straight line with label ``Landauer''
corresponds to the Landauer's dependence (\ref{N-res1})
with $|r|=g$.
}
\label{fig-chibig}
  \end{center}
\end{figure}

However, removing the term proportional to $\cos(\Psi_n)$ from
the right-hand side of Eq. (\ref{recrho2}) is not well justified operation,
because averaging over fluctuations of phase $\Psi_n$ must be performed
in the final expression for ${\cal N}_n$, and not at some intermediate
steps (as soon as $\langle f(x)\rangle \neq f(\langle x\rangle)$).
Moreover, if the random variable $\Psi_n$ is uniformly distributed
over some interval $(-\chi,\chi)$, then the average value
$\langle\cos(\Psi_n)\rangle = \sin(\chi)/\chi$
is equal (or close) to zero either if the ratio $\chi/\pi$
is an integer or if $\chi \gg 1$. In Fig. \ref{fig-chibig} we show
the results of numerical calculations of
the dependence of the natural logarithm $\ln\langle{\cal N}_n\rangle$
of the average number of photons created after $n$ pulses with
completely random phase $\chi_k$ (defined by Eq. (\ref{stoch})),
uniformly distributed over some large interval $(-\chi,\chi)$, for
 $\chi > 1$ and $\theta=0$ (i.e., in the absence of systematic shifts).
For each choice of $\chi$, averaging was performed over $20$ realizations.
%\end{multicols}
%\begin{multicols}{2}
Qualitatively, we see an agreement with Landauer's predictions: the growth
of the number of photons (or equivalent resistance) is close to exponential,
although the necessary
number of pulses is much greater than in the case of $\chi < 1$
illustrated in the Figs.~\ref{fig-chi}-\ref{fig-crit}.
However, the results of numerical experiments are close to the asymptotical
Landauer's dependence (\ref{N-res1}),
$\ln{\cal N}_n = 2n|r|^2 - \ln 2$
(given by the dashed straight line labeled as ``Landauer''),
only for integral
ratios $m=\chi/\pi$ (take into account the logarithmic scale),
whereas the inclinations
of curves corresponding
to half-integral values of the ratio $\chi/\pi=m+1/2$ (when
$|\langle\cos(\chi_n)\rangle| = [\pi(m+1/2)]^{-1}$)
are much bigger for small and moderate values of $m$, approximating
Landauer's limit only when $m>100$.

The results of this section are in qualitative agreement with other studies,
e.g.,
\cite{Petr03} (where {\em quasiperiodic\/} motions of the
boundaries were considered),
\cite{Long00} (parametric resonance in periodic paraxial optical systems),
or
\cite{Fer,Cros69,Mol70} (devoted to quantum oscillators with
fluctuating frequencies).
A possibility of generating {\em coherent phonons\/}
in solids with time-dependent lattice strains was studied in \cite{FerB}.
Mathematically, this problem is reduced to quantum oscillator with
time-dependent (fluctuating) frequency. Another mechanism of producing
coherent phonons by femtosecond laser pulses, which is mathematically
equivalent to the
problem of quantum oscillator with time-dependent {\em force\/},
was considered recently in \cite{Loz03}.

\section{Conclusion}

We have considered a simple model of photon generation from vacuum
due to nonstationary (dynamical) Casimir effect
in a
cavity with nonequidistant spectrum of eigenfrequencies, which is
reduced to the problem of quantum harmonic oscillator with time-dependent
frequency. We have shown that
periodic displacements of the (effective) boundary must result in
exponential growth of the mean number of quanta (photons) in the
selected mode under certain resonance conditions.
The oscillations of the boundary need not to
be monochromatic, but they must be close to periodic.
However, although the period of motion of the boundary, $T_b$,
should be adjusted to the period of field oscillations, $T_f$,
and to some phase, depending on the concrete form of pulse,
$T_b$ may be much greater than $T_f$. This result
seems to be very important from the point of view of facilitating
performing experiments.
The concrete
form of the trajectory of the wall enters the final result only
through effective reflection and transmission coefficients from
some ``barriers'' in the time dependence of the eigenfrequency.

We have studied the influence of deviations from the strict resonance,
caused by systematic and random perturbations of the optimal
resonance trajectory of the boundary. It appears that amplitude
perturbations are less important than the phase ones. In turn,
among the phase perturbations, the most dangerous for preserving
the regime of photon generation are systematic deviations,
which should not exceed rather low level of the order of the
frequency modulation depth.
On the contrary, random phase fluctuations can have much greater
amplitude (even of the order of unity for zero systematic deviations)
without qualitative changes in the behaviour of the system.
This circumstance also seems to be important for planning future experiments,
showing that some requirements might be not so rigid as
one could suppose, thus diminishing the number of technical problems
to be solved.

The results obtained can be easily generalized to the case when the
initial state of the field mode was not vacuum, but a thermal state.
Actually, one should simply multiply the right-hand side of Eq. (\ref{Ngen})
by the factor
$1+2\langle n_{th}\rangle$, where $\langle n_{th}\rangle$ is the mean
number of thermal photons in the initial state
\cite{DKN93a,Jing00,AD00,Pluntemp}.

However, the presented study should be considered only as a qualitative
model, because many important things have not been taken into account.
For example, we considered the excitation of a single selected mode,
neglecting its possible interactions with other modes. Such an approach
seems to be justified for cavities with nonequidistant spectra, when
the boundary performs {\em harmonic\/} oscillations at the resonance
frequency \cite{D95}. But the spectrum of anharmonic oscillations
contains many frequencies. On the one hand, this fact, perhaps, explains
why photons can be generated even if
the period of displacements of the boundary is greater than the
period of the field oscillations. On the other hand, in the presence
of many frequencies some other modes can be excited, too. This question
needs a detailed investigation. Also, the effects of polarization of true
electromagnetic field (not its scalar model)
could be important, as was shown in other examples \cite{Croc2,NetoMa96}.
And, of course, the problem of consistent account of losses in cavities
with moving nonideal mirrors must be solved (perhaps, following the lines
generalizing the approaches of Refs. \cite{D98a,Plun02,Saito}).

\section*{Acknowledgement}
The authors are grateful to the Brazilian agency CNPq for the support.

%\newpage

%\end{multicols}
%\vfill

\end{document}